\documentclass{article}

\usepackage{makeidx}  
\usepackage{amsmath}
\usepackage{amssymb}
\usepackage[boxed,algoruled,linesnumbered]{algorithm2e}
\usepackage[numbers]{natbib}

\usepackage{graphicx}
\usepackage{caption}
\usepackage{subcaption}

\usepackage{todonotes}

\newtheorem{example}{Example}

\newcommand{\mli}[1]{\mathit{#1}}

\begin{document}
\pagestyle{headings}  

\title{Monte Carlo Techniques for Approximating the Myerson Value---Theoretical and Empirical Analysis}
\author{M. K. Tarkowski$^{1}$, S. Matejczyk$^{2}$, T. P. Michalak$^{3}$, and M. Wooldridge$^{1}$\\
$^{1}$ Department of Computer Science, University of Oxford, UK\\
$^{2}$ Institute of Computer Science, Polish Academy of Sciences, Poland\\
$^{3}$ Institute of Informatics, University of Warsaw, Poland}

%
%

\date{}
\maketitle              

\begin{abstract}
Myerson \cite{Myerson:1977} first introduced graph-restricted games in order to model the interaction of cooperative players with an underlying communication network. A dedicated solution concept---the Myerson value---is perhaps the most important normative solution concept for cooperative games on graphs. Unfortunately, its computation is computationally challenging. In particular, although exact algorithms have been proposed \cite{Skibski:et:al:2014},
they must traverse all connected coalitions of the graph of which there may be exponentially many.

In this paper, we consider the issue of approximating the Myerson value for arbitrary graphs and characteristic functions. While Monte Carlo approximations have been proposed for the related concept of the Shapley value \cite{Bachrach:2010,Castro:2009,Maleki:2013}, their suitability for the Myerson value has not been studied. Given this, we evaluate and compare (both theoretically and empiraclly) three Monte Carlo sampling methods for the Myerson value: conventional method of sampling permutations; a new, hybrid algorithm that combines exact computations and sampling; and sampling of connected coalitions. We find that our hybrid algorithm performs very well and also significantly improves on the conventional methods. 
\end{abstract}
\section{Introduction}
In various cooperative settings, the assumption that all players can cooperate seems unnatural. For instance, agents in multi-agent system may have limited communication capabilities that naturally restrict coalitions that they are able to join. They may also exist various other tangible or intangible constrains such as negative experience, distrust, lack of social relationship or simply geographical distance.

To model such situation, Myerson \cite{Myerson:1977} introduced graph-restricted games in which it is assumed that only connected players can cooperate. Although players in a coalition do not have to be directly connected to communicate, there must exist a path between them using only nodes in the coalition. In essence, the value of a disconnected coalition becomes the sum of the values of its connected components.
Moreover, Myerson introduced a payoff division scheme, now called the Myerson value, which he uniquely defined by a set of two desirable axioms. It is the equivalent of the well-known concept of the Shapley value \cite{Shapley53} in graph-restricted games and it has various application. In particular, it is one of the key concepts in the literature on game-theoretic centrality in networks \cite{Gomez:et:al:2003,skibski2019attachment,Michalak:et:al:2013,tarkowski2018efficient,skibski2018axiomatic} (with applications, for instance, to covert networks \cite{Michalak:et:al:2013b,michalak2015defeating}, weighted voting gamaes \cite{skibski2015pseudo,skibski2017algorithm} or measuring social capital~\cite{michalak2015new}).

The state-of-the-art algorithm to compute the Myerson value is due to Skibski et al.~\cite{skibski2019enumerating}. However, it traverses all connected coalitions of the graph of which there may be exponentially many. Thus, exact computations are infeasible in many fields of interest, such as social network analysis (where various methods are based on the Myerson value \cite{Gomez:et:al:2003}). As for approximate algorithms, the Monte Carlo sampling methods have been studied for the Shapley value \cite{Bachrach:2010,Castro:2009,Maleki:2013} in conventional coalitional games. However, their suitability for the Myerson value has not been established.

Against this background, in this paper, we consider the issue of approximating the Myerson value for arbitrary graphs and characteristic functions. Algorithm \ref{algorithm:1} is a variation on a well-known sampling method for the Shapley value. Our new, hybrid Algorithm \ref{algorithm:hybrid} combines both exact computations and sampling. Our algorithm \ref{algorithm:connected} samples coalitions, but only performs calculations for connected ones. We analyse all three algorithms empirically and establish probably-approximately correct bounds on the number of necessary samples. It turns out that our Algorithm \ref{algorithm:hybrid} achieves the best results and even improves on the literature for approximating the Shapley value.

\section{Preliminaries}
We will begin by introducing some concepts that will be necessary for the rest of the paper. We begin with the graph theoretic concepts, and then move on to cooperative game theory.

\vspace{0.2cm} \noindent\textbf{Graph Theory:} For the purposes of this paper, we will assume that a network is an undirected graph. A graph, $G = (V,E)$, consists of a set of nodes, $V$, and a set of edges, $E \subseteq V \times V$. Since we are concerned with undirected graphs, we will assume that if $(u,v) \in E$, then $(v,u) \in E$. Let $V' \subseteq V$. We will say that $G(V')$ is the subgraph of $G$ induced by the set of nodes $V'$ if and only if $G(V') = (V', E')$ such that $E' = E \cap V' \times V'$. A path between the nodes $v_1$ and $v_k$ is an ordered set of nodes $(v_1, v_2, \ldots, v_k)$ such that for all $1 \leq i  \leq k - 1$ we have $(v_i, v_{i+1}) \in E$. We say that a graph is connected if for all pairs of nodes $s, t \in V$ there exists a path between them. By $\mathcal{C}(G)$ we denote the set of \textit{connected components} of the graph $G$. That is, $\mathcal{C}(G) = \{ V' : V' \subseteq V, G(V') \text{ is connected}, \forall_{v \in V} G(V' \cup \{v\}) \text{ is not connected } \}$. By $\mathcal{K}(V') = \mathcal{C}(G(V'))$ we denote the set of connected components of the graph induced by $V' \subseteq V$. Finally, by $N(v) = \{u: (v,u) \in E\}$ we denote the set of neighbours of $v \in V$ and by $N(V') = \{u: \exists_{v \in V'} (v,u) \in E \text{ and } u \not\in C\}$ the set of neighbours of $V' \subseteq V$.

\vspace{0.2cm} \noindent\textbf{Cooperative Game Theory:} A cooperative game consists of a set of players, $P$, and a characteristic function, $\nu: 2^P \rightarrow \mathbb{R}$, where $\nu(\emptyset) = 0$. In other words, a cooperative game consists of players and a valuation function of all the \textit{coalitions} (i.e., subsets) of the set of players. If $C \subset P$ and $i \not\in C$, then by $MC(C, i) = \nu(C \cup \{i\}) - \nu(C)$ we will denote the marginal contribution of player $i$ to the coalition $C$. We will now introduce a few common types of games from the literature (for example, they can be found in \cite{Shapley:1971}):

\begin{itemize}
\item \textbf{Super-additive:} $\nu$ is superadditive if and only if for any $S, T \subseteq V$ such that $S \cap T = \emptyset$, we have $\nu(S \cup T) \geq \nu(S) + \nu(T)$. That is, the union of any two disjoint coalitions is worth at least as much as the sum of their values.
\item \textbf{Submodular:} $\nu$ is submodular if and only if $\nu(S \cup T) + \nu(S \cap T) \leq \nu(S) + \nu(T)$. This intuitively captures the inverse of the snowballing effect: it is less and less worthwile to join larger coalitions.
\item \textbf{Symmetric:} $\nu$ is symmetric if it only depends on the size of the coalition. In other words, $\nu(C) = f(|C|)$ for some $f: \mathbb{N} \rightarrow \mathbb{R}$.
\end{itemize}
Given this setting, a fundamental question is that---assuming that the \textit{grand coalition} (i.e., coalition of all the players) forms---how do we divide the value of the coalition among its members? Such methods of handing out the \textit{payoffs} to players are referred to as \textit{payoff division schemes} or \textit{values of the game}. Shapley \cite{Shapley:1953} famously introduced a value, now known as the Shapley value, as one such division scheme. 

\begin{itemize}
\item \textbf{Efficiency}---The whole value of the grand coalition is distributed. In other words $\sum_{i \in P} SV_i(\nu) = \nu(P)$.
\item \textbf{Symmetry}---The payoffs of agents do not depend on their identity.
\item \textbf{Null Player}---A player who does not contribute to any coalition will have no payoff. For any player $i$, if for all $C$ we have $MC(C, i) = 0$, then $SV_i(\nu) = 0$.
\item \textbf{Additivity}---For any two games with the same agent set, $(P, \nu_1)$ and $(P, \nu_2)$, the payoff from the sum of the two games will be the sum of payoffs from each game separately. That is, $SV_i((\nu_1 + \nu_2)) = SV_i(\nu_1) + SV_i(\nu_2)$ where $(\nu_1 + \nu_2)(C)$ is equal to $(\nu_1(C) + \nu_2(C))$.
\end{itemize}
Shapley also showed that the axioms are equivalent to the following formula:
\begin{equation} \label{equation:shapley:permutation}
SV_i(\nu) = \sum_{\pi \in \Pi(P)} \frac{MC(C_\pi(i), i)}{|P|!},
\end{equation}
where $SV_i(\nu)$ denotes the Shapley value of player $i$ in the game with characteristic function $\nu$, $\Pi(P)$ is the set of permutations of the set $P$ and $C_\pi(i)$ is the set of all players preceding $i$ in the permutation $\pi$. This is also equivalent to the following:
\begin{equation} \label{equation:shapley:subset}
SV_i(\nu) = \sum_{C \in (P \setminus \{i\}} \frac{|C|!(|P| - |C| - 1)!}{|P|!} MC(C, i).
\end{equation}
The Shapley value is the unique division scheme that satisfies not only the one above but also other various intuitive axiomatizations (see the recent work by Maschler et al. \cite{Maschler:et:el:2013} for a sample overview).

\vspace{0.2cm}\noindent\textbf{Combining Networks and Cooperative Games:} In this paper, we are focused on a model due to Myerson that combines both a network setting and a cooperative game. The model consists of a network, and a cooperative game on the set of nodes $(G = (V,E), \nu: 2^V \rightarrow \mathbb{R})$. Myerson's reasoning was that players who are not connected should not be allowed to cooperate. Therefore, the value of a coalition that is disconnected should be the sum of its connected components.\footnote{There also exist other methods of restricting cooperative games. For example, the value of a disconnected coalition can be set equal to zero.} In other words, Myerson introduced \textit{graph-restricted games}, i.e., the procedure of restricting games according to some graph structure. We will denote the characteristic function $\nu$, when it is restricted by the graph $G$, as $\nu_G$:
\begin{equation}
\nu_G(C) = \sum_{K \in \mathcal{K}(C)} \nu(C)
\end{equation}
Most importantly, Myerson introduced the Myerson value as a division scheme for this setting. We will denote the Myerson value of node $v$ in graph $G$ with characteristic function $\nu$ by $MV_v(G;\nu)$. The Myerson value is the unique payoff division scheme that satisfies the following two axioms:
\begin{itemize}
\item \textbf{Efficiency in Connected Components:} The whole value of the grand coalition according to the restricted function is distributed. In effect, we assume that the coalitions of connected components have formed, rather than the grand coalition. Formally, $\sum_{v \in V} MV_v(\nu) = sum_{C \in \mathcal{C}(G)} \nu(C)$
\item \textbf{Fairness:} If we add an edge, $(s,t)$, to $G$ (denoted by $G \cup (s,t)$), then the Myerson value of nodes $s$ and $t$ changes by the same amount. Formally: 
\[MV_s(G \cup (s,t); \nu) - MV_s(G; \nu) = MV_t(G \cup (s,t); \nu) - MV_t(G; \nu).\]
\end{itemize}
Myerson showed, that the value he defined is equivalent to the Shapley value of the graph-restricted game:
\begin{equation}
MV_v(G; \nu) = SV_v(\nu_G).
\end{equation}

In the next section, we consider the issue of approximating the Myerson value.

\section{Approximation Algorithms}
In this section, we present three algorithms for approximating the Myerson value. All of them rely on Monte Carlo sampling. In essence, we rewrite the Myerson value as an average of a certain set of values that are based on the marginal contributions of players. Next, we approximate $MV_i$ by sampling certain elements from this set, and taking the average of the samples. Such computational techniques are necessary, since the best algorithms for computing the Myerson value exactly must iterate over all connected subgraphs of a graph, of which there may be exponentially many. This makes exact computation infeasible for many networks even when they are of moderate size.

\vspace{0.2cm}\noindent\textbf{Sampling Permutations:} Algorithm \ref{algorithm:1} is a variation on the well known Monte Carlo sampling of permutations for the Shapley value. The key idea is that the Shapley value is an average of the marginal contributions of a player to all permutations (Equation \ref{equation:shapley:permutation}). Sampling permutations, then, we can take the average of our samples in order to estimate the Shapley value. That is, for $m$ samples, our estimate of the Shapley value will be $\widehat{SV}_i(\nu) = \frac{1}{m} \sum_{i = 1}^m MC(C_{\pi}(i),i)$, where $\pi$ is a random variable representing a random permutation of the set $V$ with a uniform probability distribution. Since the Myerson value of $\nu$ is simply the Shapley value of $\nu_G$, we can use the same method for estimating $SV_i(\nu_G)$.

Line 4 is the main loop of the algorithm, where samples are drawn and added to the estimate. Note that we do not explicitly sample permutations, but instead directly sample the set $C_{\pi}(i)$). Sampling coalition size with uniform probability (line 6), and then a coalition of that size with uniform probability (line 7) mimics sampling permutations. This has the added advantage that sampling coalitions is computationally simpler. Moreover, the algorithm can be easily modified for computing any semivalue\footnote{This is beyond the scope of this article. For an overview on semivalues see \cite{carreras:2002}} by modifying the distribution in line \ref{alg:line:dist}.

Line \ref{algorithm:permutations:line:s} and \ref{algorithm:permutations:line:swap} may seem unintuitive, but are necessary in order to use a sample not for just one, but for all nodes. Were we sampling for just one node, $v$, we could sample from the set $V \setminus v$ (since a node can only make a contribution to a set of which it is not a part). What we do instead, is sample from a set that is missing one node, $s$, and then decide which node that is afterwards by swapping the $v$ with $s$. This is done in line \ref{algorithm:permutations:line:swap}, where $S_{v \leftrightarrow s}$ is the set of elements from $S$ with the difference that if $v \in S$, then $v \not\in S_{v \leftrightarrow s}$ and $s \in S_{v \leftrightarrow s}$ (i.e. the elements are swapped). If $s = v$, we assume $S = S_{v \leftrightarrow s}$. Finally, in line 10 we add the marginal contribution, and in line 12 take the average over all marginal contributions.

\RestyleAlgo{ruled}
\begin{algorithm}
\SetAlgoVlined
\LinesNumbered
\KwIn{Graph: $G=(V,E)$, Function: $\nu: \mathcal{C}(G) \rightarrow \mathbb{R}$, Number Samples: $m$}
 $MV \gets$ array of size $|V|$ filled with $0$\;
 $s \gets$ arbitrary node in $V$\;\label{algorithm:permutations:line:s}
 \For{$i \gets 1$ to $m$}{
   //sample an integer from the set $[0, |V|-1]$ with uniform probability\;
   $k \gets UNIFORM\_DIST(0,|V| - 1)$\; \label{alg:line:dist}
   $S \gets$ random coalition of size $k$ from the set $V \setminus s$\;
   \For{$v \in V$}{
      $C \gets S_{v \leftrightarrow s}$\;\label{algorithm:permutations:line:swap}
   	  $MV[v] \gets MV[v] + \nu_G(C \cup \{v\}) - \nu_G(C)$\;
   }
 }
 \For{$v \in V$}{
	$MV[v] \gets \frac{MV[v]}{m}$\;
 }
  
 Output: $MV$\;
 \caption{Sampling permutations for computing the Myerson value}
 \label{algorithm:1}
\end{algorithm}

\noindent\textbf{Hybrid Algorithm:} The idea behind our new, hybrid Algorithm \ref{algorithm:hybrid}, is to perform exact computations for permutations, where the node $v$ is near the beginning or the end of the permutation. In particular, we will consider exact calculations when $k \leq \pi(i)$ or $\pi(i) \leq |V| - k$. This is motivated by the fact that for any $j$, there is the same number of coalitions of size $j$ as there is of size $|V| - j$. For example when $k=0$, we consider when $i$ is the first or the last element in any permutation. Note that there are $2(|V| - 1)!$ such permutations, but only two coalitions to which $i$ makes a contribution: $\emptyset$, or $V \setminus \{i\}$.
Similarly, there are $\binom{|V|}{k} = \binom{|V|}{|V| - k}$ coalitions to which $i$ makes a contribution, when $\pi(i) = k$ or $\pi(i) = |V| - k$. We will use this fact in order to simplify computation.
We will discuss the advantage of this approach in Section \ref{section:theory}.

We will now rewrite Equation \ref{equation:shapley:permutation} according to the position of player $i$ and combine this with \ref{equation:shapley:subset} for the exact computations:
\begin{align}
SV_i = \hspace{-1cm} &\sum_{\substack{C \in (V \setminus \{i\}) \\ |C| \leq k \text{ or } |C| \geq |V| - k - 1}} \hspace{-0.8cm} \frac{|C|!(|V| - |C| - 1)!}{|V|!}MC(C_{\pi}(i), i) + \hspace{-0.5cm}  \sum_{\substack{\pi \in \Pi(V)\\ k < \pi(i) < |V|- k - 1}} \hspace{-0.3cm} \frac{MC(C_{\pi}(i), i)}{|V|!} \nonumber
\end{align}

\vspace{-0.5cm}

\RestyleAlgo{ruled}
\begin{algorithm}
\SetAlgoVlined
\LinesNumbered
\KwIn{Graph: $G=(V,E)$, Function: $\nu: \mathcal{C}(G) \rightarrow \mathbb{R}$, Number Samples: $m$, Number exact computations: $Ex$}
 $MV \gets$ array of size $|V|$ filled with $0$\;
 $s \gets$ some node in $V$\;
 \For{$i \gets 1$ to $m$}{
    //uniform distribution from the set $[Ex + 1,|V| - Ex - 2]$\;
   $k \gets UNIFORM\_DIST(Ex + 1,|V| - Ex - 2)$\;
   $S \gets$ random coalition of size $k$ from the set $V \setminus s$\;
   \For{$v \in V$}{
      $C \gets S_{v \leftrightarrow s}$\;
   	  $MV[v] \gets MV[v] + \nu_G(C \cup \{v\}) - \nu_G(C)$\;
   }
 }
 \For{$v \in V$}{
 	$MV[v] \gets \bigg(\frac{|V| - 2Ex - 2}{|V|}\bigg) 
 	\frac{MV[v]}{m}$\;
 }
 
 $E \gets \{C : C \subset V \text{ and } |C| \leq Ex\}$\;
 \For{$C \in E$}{
   $val\_C \gets \nu_G(C)$\;
   \For{$v \in (V \setminus C)$}{
   	  $MV[v] \gets MV[v] + \frac{|C|!(|V| - |C| - 1)!}{|V|!} (\nu_G(C \cup \{v\}) - val\_C)$\;
   }
   $C \gets V \setminus C$\; \label{algorithm:exact:complement}
   $val\_C \gets \nu_G(C)$\;
   \For{$v \in C$}{
   	  $MV[v] \gets MV[v] + \frac{(|C|-1)!(|V| - |C|)!}{|V|!} (val\_C - \nu_G(C \setminus \{v\})$\;
   }
 }
  
 Output: $MV$\;
 \caption{Our algorithm: sampling permutations \& exact computations to approximate the Myerson value}
  \label{algorithm:hybrid}
\end{algorithm}

\noindent We will estimate the second term of Equation \ref{equation:hybrid} by rewriting it as an average over the  $\frac{(|V| - 2|k| - 2)|V|!}{|V|}$ permutations $\pi$ such that $\pi \in \Pi(V), k < \pi(i) < |V|- k$:
\begin{align}\label{equation:hybrid}
\frac{|V|}{(|V| - 2|k| - 2)|V|!}\sum_{\substack{\pi \in \Pi(V)\\ k < \pi(i) < |V|- k - 1}} \bigg(\frac{|V| - 2|k| - 2}{|V|}\bigg)MC(C_{\pi}(i), i)),
\end{align}

\noindent leading us to the estimate with random permutation $\pi$:

\begin{align*}
\frac{1}{m}\sum_{i = 1}^m \bigg(\frac{|V| - 2|k| - 2}{|V|}\bigg) MC(C_{\pi}(i), i),
\end{align*}

\vspace{0.2cm}\noindent\textbf{Sampling Connected Coalitions:} Let $\mathcal{C}(i) = \{C : i \in C \text{ and } \{C\} = \mathcal{K}(C)\}$ be the number of coalitions such that $i \in C$ and $C$ is connected and let $\mathcal{N}(i) = \{C : i \in N(C) \text{ and } \{C\} = \mathcal{K}(C)\}$ be the number of coalitions such that $i \in N(C)$ and $C$ is connected. We will now consider the following formula for the Myerson value (due to \cite{Skibski:et:al:2014}):
\begin{equation}
MV_i(\nu) = MV^+_i(\nu) - MV^-_i(\nu),
\end{equation}
where:
\begin{align}
\label{equation:myerson:positive}
&MV^+_i(\nu) = \sum_{C \in \mathcal{C}(i)} \frac{(|C| - 1)!|N(C)|!}{(|C| + |N(C)|)!} \nu(C),\nonumber\\
&MV^-_i(\nu) = \sum_{C \in \mathcal{N}(i)} \frac{(|C|)!|N(C) - 1|!}{(|C| + |N(C)|)!} \nu(C).
\end{align}
We can also rewrite these equations as averages over a set of elements:
\begin{align*}
&MV^+_i(\nu) = \frac{1}{|\mathcal{C}(i) |}\sum_{C \in \mathcal{C}(i)} |\mathcal{C}(i)| \frac{(|C| - 1)!|N(C)|!}{(|C| + |N(C)|)!} \nu(C),\\
&MV^-_i(\nu) = - \frac{1}{|\mathcal{N}(i)|} \sum_{C \in \mathcal{N}(i)} |\mathcal{N}(i)| \frac{(|C|)!|N(C) - 1|!}{(|C| + |N(C)|)!} \nu(C).
\end{align*}
Drawing a random connected coalition from the set of connected coalitions, we can attempt to estimate $MV^-_i(\nu)$ and $MV^+_i(\nu)$, however we do not know the values of $|\mathcal{C}(i)|$ and $|\mathcal{N}(i)|$. If we sample random coalitions (not only those that are connected) with uniform probability, we will note that we can estimate these values. Let $m$ be the number of samples, and $m(C_i)$ be the number of samples such that they fall in the set $\mathcal{C}(i)$, and $m(N_i)$ be the number of samples such that they fall into the set $\mathcal{N}(i)$. Notice that $\frac{m(C_i)}{m}$ is an estimate of the percentage of all coalitions that fall into the set $\mathcal{C}(i)$ (and similarly for $\mathcal{N}(i)$). The total number of non-empty coalitions is $2^{|V|} - 1$, and hence $(2^{|V|} - 1) \frac{m(C_i)}{m}$ gives us an estimate of $|\mathcal{C}(i)|$. We can then use the following formula as an estimate of $MV^+_i(\nu)$ (and similarly for $MV^-_i(\nu)$):
\begin{align}
\widehat{MV}^+_i(\nu) &= \frac{1}{m(C_i)}\sum_{C \in \mathcal{C}(i)} (2^{|V|} - 1) \frac{m(C_i)}{m} \frac{(|C| - 1)!|N(C)|!}{(|C| + |N(C)|)!} \nu(C)\nonumber\\
&= \frac{1}{m} \sum_{C = \mathcal{K}(C), i \in C}  (2^{|V|} - 1) \frac{(|C| - 1)!|N(C)|!}{(|C| + |N(C)|)!} \nu(C) \label{equation:myerson:coalition:sampling}
\end{align}
Note how $m(C_i)$ disappears from the equation. An alternative way of looking at things is to consider Equations \ref{equation:myerson:positive} as an average over all non-empty coalitions (of which there are $2^{|V|} - 1$), where disconnected coalitions have value $0$. This serves as an equivalent interpretation of Equation \ref{equation:myerson:coalition:sampling}, since the average is now over $m$ sampled coalitions.

\RestyleAlgo{ruled}
\begin{algorithm}
\SetAlgoVlined
\LinesNumbered
\KwIn{Graph: $G=(V,E)$, Function: $\nu: 2^{|V|} \rightarrow \mathbb{R}$, Number Samples: $m$}
 $MV \gets$ array of size $|V|$ filled with $0$\;
 $MV^- \gets$ array of size $|V|$ filled with $0$\;
 $MV^+ \gets$ array of size $|V|$ filled with $0$\;
 \For{$i \gets 1$ to $m$}{
   $C \gets$ random, non-empty coalition\; 
   \If{$C$ is connected}{
   		\For{$v \in C$}{
   			$MV^+[v] \gets MV^+[v] + \frac{(|C|-1)!|N(C)|!}{(|C| + |N(C)|)!} \nu(C)$\;
   		}
   		\For{$v \in N(C)$}{
   			$MV^-[v] \gets MV^-[v] \frac{|C|!(|N(C)| - 1)!}{(|C| + |N(C)|)!} \nu(C)$\;
   		}
   }
 }
 \For{$v \in V$}{
    $MV[v] \gets \frac{2^{|V|} - 1}{m} (MV^+[v] - MV^-[v])$\;
    }
 Output: $MV$\;
 \caption{Sampling connected coalitions for computing the Myerson value}
 \label{algorithm:connected}
\end{algorithm}

\section{Theoretical Analysis} \label{section:theory}
As has become the standard in the literature \cite{Maleki:2013,Bachrach:2010}, we will use one of Hoeffding's theorems in order to achieve a number of samples that will result in a probably approximately correct approximation. That is, we are looking for the number of samples, $m$, that will result in the following: $Pr(|MV_i(\nu) - \widehat{MV}_i(\nu)| \geq \epsilon) \leq \delta$.

Hoeffding's Theorem 2 \cite{Hoeffding:1963} will allow us to bound the error within some probability by knowing the minimum and maximum value of the marginal contribution of nodes. The theorem implies\footnote{Similarly to \cite{Maleki:2013}, we have provided a less general version, more suited to our context.} that if $S = \sum_{i = 1}^m X$, where $X$ is a random variable with range (i.e., the difference between the maximum and minimum values) $r$, then for $\epsilon > 0$ we have:
\vspace{-0.4cm}
\begin{equation}
Pr(|\phi - \hat{\phi}| \geq \epsilon) \leq 2\exp{\bigg(-\frac{2m^2\epsilon^2}{mr^2}\bigg)}.
\vspace{-0.1cm}
\end{equation}
In our context, $X$ will be based on the marginal contributions of a player (with some modifications, depending on the sampling method). This leads to the following inequality:
\vspace{-0.2cm}
\begin{equation}
2\exp({-\frac{2m\epsilon^2}{r^2}}) \leq \delta \hspace{0.3cm} \Rightarrow \hspace{0.3cm}
-\frac{2m\epsilon^2}{r^2} \leq ln(\frac{\delta}{2}) \hspace{0.3cm} \Rightarrow \hspace{0.3cm}
m \geq -\frac{ln(\frac{\delta}{2})r^2}{2\epsilon^2}.
\vspace{-0.2cm}
\end{equation}

Note, however, that $MC(v,C)$ is based on $\nu_G$, and not $\nu$, so simply knowing the range of the marginal contributions for $\nu$ will not suffice. In fact, the range for $\nu_G$ can strongly depend on the graph. Let us look at the following example:

\begin{figure}
\center
\scalebox{0.8}{
\begin{tikzpicture}[every node/.style={draw=black,circle}]
  \node (1)  {$v_1$};
  \node (2)  [right=1cm of 1] {$v_2$};
  \node (3)  [left=1cm of 1] {$v_3$};
  \node (4)  [above=1cm of 1] {$v_4$};
  \node (5)  [below=1cm of 1] {$v_5$};
  \draw[-] (1) edge (2);
  \draw[-] (1) edge (3);
  \draw[-] (1) edge (4);
  \draw[-] (1) edge (5);

\end{tikzpicture}
}
\caption{A sample star network}\label{figure:star:network}
\end{figure}
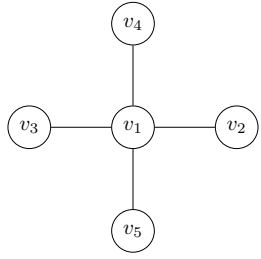

\begin{example}
Consider Figure \ref{figure:star:network} and the characteristic function $\nu(C) = |C|^2$. We see that the minimum marginal contribution of a node is $1$, whereas the maximum contribution is $5^2 - 4^2 = 9$. However, what if we consider $\nu_G$? The contribution of node $v_1$ to $\{v_2, v_3, v_4, v_5\}$ is equal to $5^2 - 4 = 21$. Because it is central in the graph, $v_1$ is able to make a much larger contribution.
\end{example}

If $\nu$ takes only positive or negative values, we can use a conservative approximation of the range: $r = \max_{C \subseteq V}(|\nu(C)|)$. In general, however, this might be a more difficult. Consider, for example, Figure \ref{figure:star:network} with the characteristic function:
\[
\nu(C) = \begin{cases}
1 & |C| = 1\\
-1 & |C| > 1\\
0 & \text{otherwise}
\end{cases}
\]

In this case, $v_1$ makes a contribution of $-5$ to $\{v_2, v_3, v_4, v_5\}$. A conservative bound for $r$ that is easy to calculate is as follows: $ r = |V|\max_{C \subseteq V}(|\nu(C)|)$.
We will now move on to each of our three sampling methods.

\noindent\textbf{Sampling Permutations:} In Algorithm \ref{algorithm:1}, our estimate for the Myerson value of node $v$ is based on the following: $\sum_{i = 1}^m \frac{MC(v,C)}{m}$, where $C$ is a random variable representing a random coalition with a probability distribution such that $Pr(|C| = k) = \frac{1}{|V|}$ for $0 \leq k \leq |V| - 1$, and coalitions of the same size are equally probable. This leads to the following number of samples:
\begin{align*}
m &\geq -\frac{ln(\frac{\delta}{2})(\frac{r}{m})^2}{2\epsilon^2} \hspace{0.5cm} \Rightarrow \hspace{0.5cm} m \geq \sqrt[3]{-\frac{ln(\frac{\delta}{2})r^2}{2\epsilon^2}}.
\end{align*}

\vspace{0.2cm}\noindent\textbf{Hybrid Algorithm:} Here,  we will see the advantage of Algorithm \ref{algorithm:hybrid}. The random variable that Algorithm \ref{algorithm:hybrid} samples over is $\sum_{i = 1}^m \big(\frac{|V| - 2Ex - 2}{|V|}\big)\frac{MC(v,C)}{m}$. Note that now $C$ is a random variable that ranges over coalitions of size $k$ such that $Ex < k < |V| - Ex - 1$. This means that especially for superadditive, subadditive, supermodular and submodular functions, the extreme values are computed exactly, and the range is further reduced. A conservative bound is $r =\max_{C \subseteq V, Ex < |C| < |V| - Ex - 1}(|\nu(C)|)$ for positive or negative valued functions, or $ r = |V|\max_{C \subseteq V, Ex < |C| < |V| - Ex - 1}(|\nu(C)|)$ for any characteristic function. The number of required samples is as follows:
\begin{align*}
m &\geq -\frac{ln(\frac{\delta}{2})((\frac{|V| - 2Ex - 2}{|V|})\frac{r}{m})^2}{2\epsilon^2}
 \hspace{0.5cm} \Rightarrow \hspace{0.5cm} m \geq \bigg(\frac{|V| - 2Ex - 2}{|V|}\bigg)^{\frac{2}{3}} \sqrt[3]{-\frac{ln(\frac{\delta}{2})r^2}{2\epsilon^2}}.
\end{align*}
In particular, notice that with just $Ex = 0$ we compute exactly the contribution of a player to only two coalitions, but decrease the required sample size by a factor of $(\frac{2}{|V|})^{\frac{2}{3}}$.

\vspace{0.2cm}\noindent\textbf{Sampling Connected Coalitions:} The random variable for Algorithm \ref{algorithm:connected} can take on one of three values: $0$, if $C$ is not connected; $\frac{2^{|V|} - 1}{m}\frac{(|C|-1)!|N(C)|!}{(|C| + |N(C)|)!} \nu(C)$, if $v \in C$; and $\frac{2^{|V|} - 1}{m}\frac{|C|!(|N(C)| - 1)!}{(|C| + |N(C)|)!} \nu(C)$, if $v \in N(C)$. The impact of $\frac{(|C| - 1)!|N(C)|!}{(|C| + |N(C)|)!}$ and $\frac{|C|!(|N(C) - 1|)!}{(|C| + |N(C)|)!}$ is unpredictable, and depends strongly on the graph. The maximum value that these fractions can take is $\frac{1}{|V|}$ and the minimum is $\alpha = \frac{(\frac{|V|}{2})!(\frac{|V|-1}{2})!}{|V|!}$, impacting the range of the random variable. We can bound the range as follows: $r = \max_{C \subseteq V}(|\nu(C)|) + \frac{\alpha}{|V|}\min_{C \subseteq V}(|\nu(C)|)$. The number of required samples, then, is:
\vspace{-0.3cm}
\begin{align*}
m \geq (2^{|V|}-1)^{\frac{2}{3}} \sqrt[3]{-\frac{ln(\frac{\delta}{2})r^2}{2\epsilon^2}}.
\vspace{-0.1cm}
\end{align*}

Although this bound may seem much worse than the other two algorithms, this analysis does not take into account the fact that computing a sample in Algorithm \ref{algorithm:connected} is significantly less expensive than in the other algorithms. In the next section, we will empirically evaluate the performance of the algorithms by limiting their running time, which will help us evaluate their performance from another perspective.

\section{Empirical Analysis}

We empirically test our algorithms on a broad spectrum of characteristic functions and graphs. The following games are considered:
\begin{itemize}
\item \textbf{Uniform random:} $\nu(C)$ is sampled from a uniform distribution over $(0, |C|)$;
\item \textbf{Uniform superadditive:} $\nu(C)$ is sampled from a uniform distribution over $(\kappa_C, \kappa_C + \mli{maxGain})$, where $\kappa_C = \max_{S,T \subseteq C, S \cup T = C, S \cap T = \emptyset}(\nu(S) + \nu(T))$ and $\mli{maxGain} = 3$ is a constant representing the degree of superadditivity;
\item \textbf{Uniform random monotone submodular:} $\nu(C)$ is sampled from a uniform distribution over $(0, \mli{maxSingleton})$ if $|C| = 1$, and from $(\lambda, \mu)$ otherwise, where $\lambda$ the minimal value that ensures monotonicity and $\mu$ is the maximal value that ensures submodularity.
\end{itemize}
\noindent We study each type of game for the following graph\footnote{results for more graphs, including Erdos-Renyi graphs, are available in the appendix}:
\begin{itemize}
\item \textbf{Preferential Attachment:} Preferential Attachment graph generation model proposed by \citet{barabasi1999emergence} with $m_0$ initially connected nodes and $m$ being the degree that a vertex has after joining. Graphs obtained using this method have a power-law degree distribution, where high degree nodes gain a high reward for their central position in the network.
\end{itemize}

\begin{figure}[h]
\centerline{
\begin{subfigure}{0.4\textwidth}
\begin{center}
\includegraphics[width=0.97\linewidth]{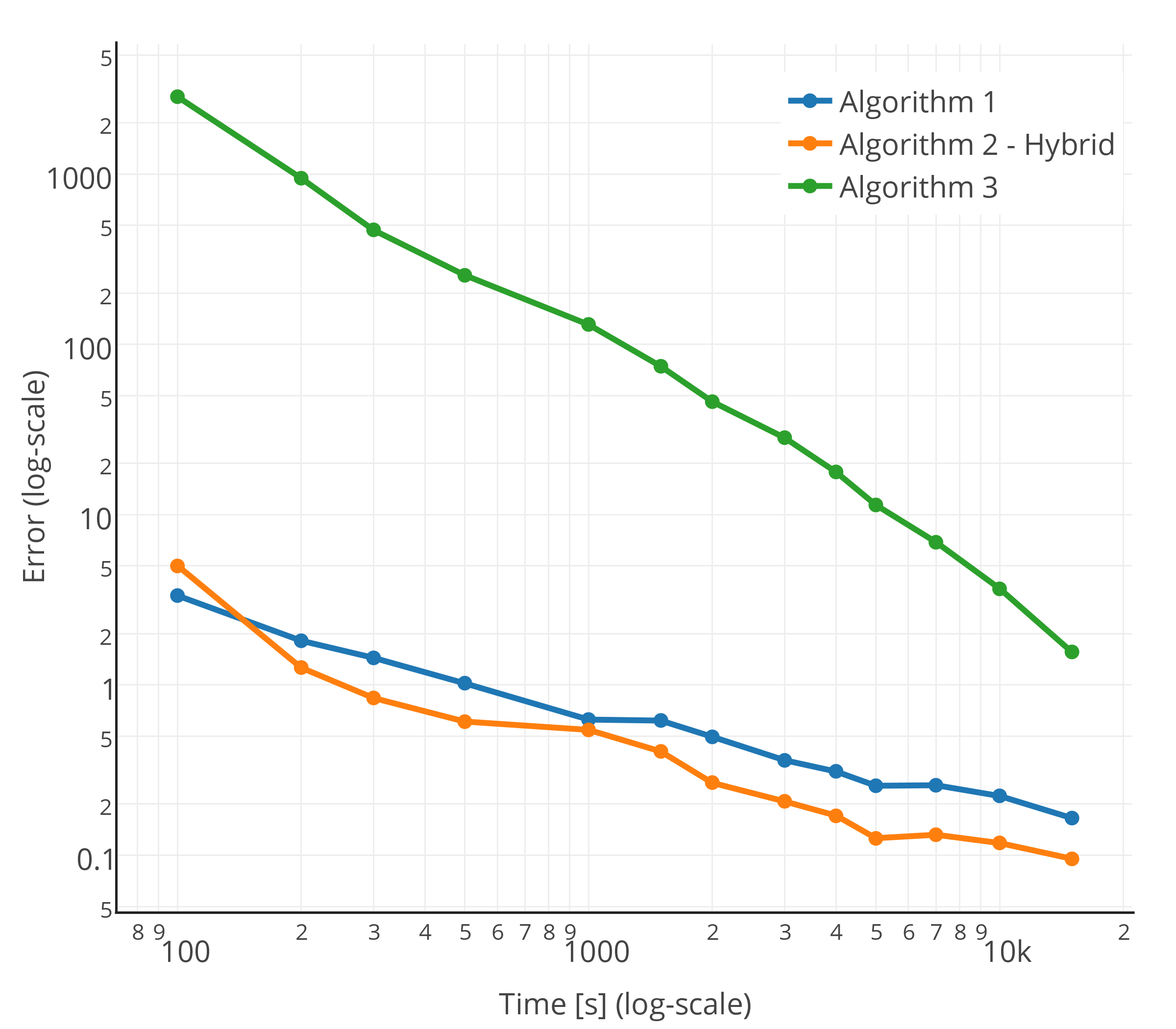}
\end{center}
\caption{Uniform random}
\end{subfigure}
\begin{subfigure}{0.4\textwidth}
\begin{center}
\includegraphics[width=0.97\linewidth]{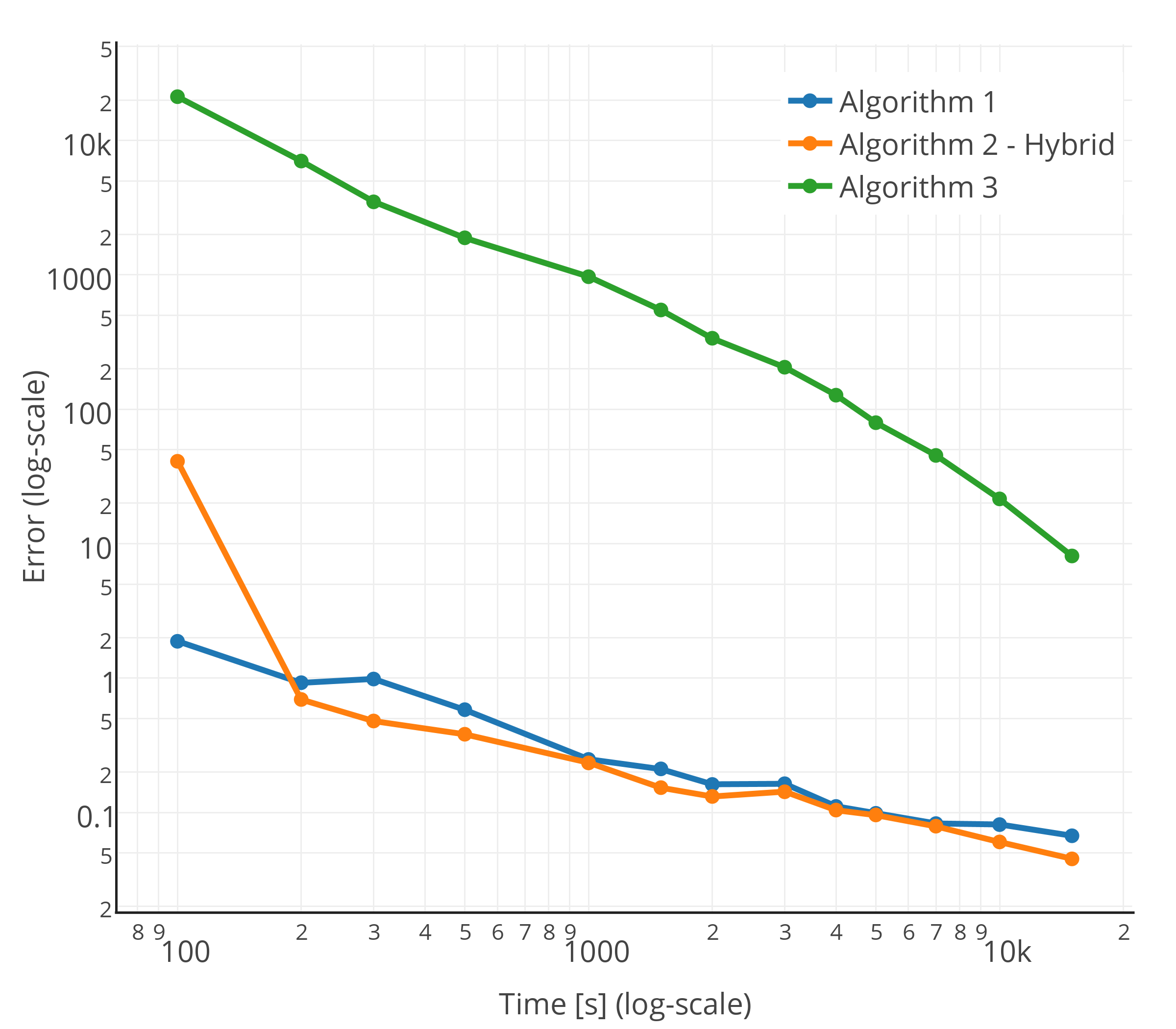}
\end{center}
\caption{Superadditive random}
\end{subfigure}
\begin{subfigure}{0.4\textwidth}
\begin{center}
\includegraphics[width=0.97\linewidth]{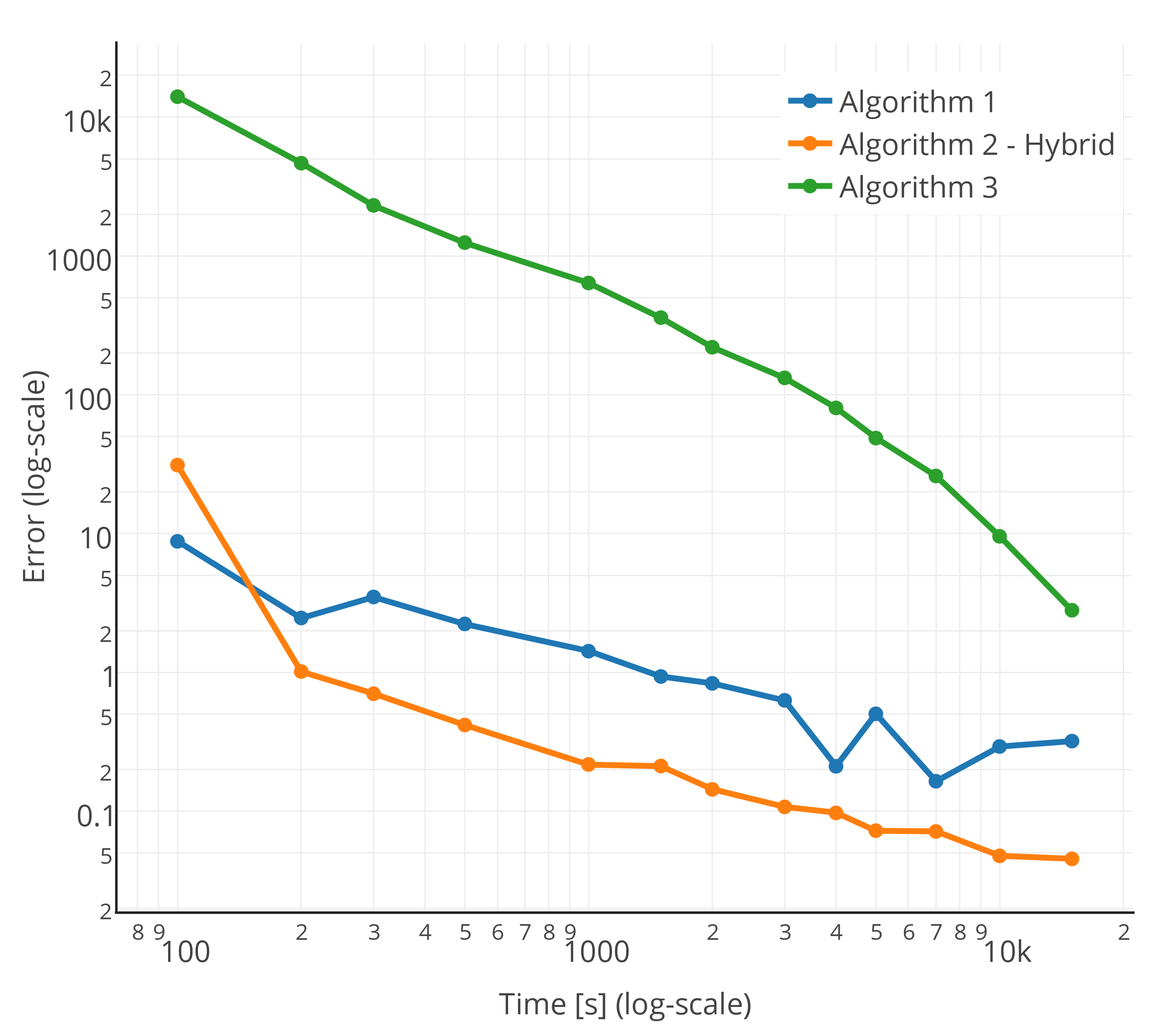}
\end{center}
\caption{Submodular random}
\end{subfigure}
}
\caption{Estimation error against the time for Preferential Attachment graph.}
\end{figure}

Due to computational restrictions, we evaluate our algorithms for $|V| = 15$. For this graph size, it is still possible to generate a characteristic function and compute the exact Myerson value of players relatively quickly. We define the error of the estimated value as the sum for all players of absolute deviations from the exact value.

\section{Discussion}
As expected, our hybrid Algorithm \ref{algorithm:hybrid} performs significantly better than simple Monte Carlo sampling in Algorithm \ref{algorithm:1}. This is reflected both in the theoretical and empirical analysis. In particular, computing the marginal contribution of players to just two coalitions (the empty coalition, and the grand coalition minus one player) already decreases the required number of samples by a factor of $\sqrt[3]{\frac{2}{|V|}}$. Although Algorithm \ref{algorithm:connected} performs worse per sample than either of the two previous algorithms, it must be kept in mind that computing samples for this algorithm is faster. Nonetheless, even when given the same running time, Algorithm \ref{algorithm:connected} cannot compete.

\section{Conclusion and Future Work}
We have analysed both empirically and theoretically three\footnote{Algorithm \ref{algorithm:1} is a variation on a well-known approximation for the Shapley value} algorithms for approximating the Myerson value. We have shown that our algorithm, Algorithm \ref{algorithm:hybrid}, achieves the best results, and improves upon the approximations of the Shapley value in the literature. Future work may include different types of sampling (i.e., sample only connected coalitions) or different approximation techniques altogether. Moreover, approximation techniques have not yet been developed for many solution concepts or for game-theoretic network centrality measures that are inspired by Myerson's graph-restricted games. In particular, the work by \cite{Amer:Gimenez:2004,Amer:Gimenez:2007} seems well suited for this.

\section*{Acknowledgements}
This paper was supported by the Polish National Science Centre grant DEC-2013/09/D/ST6/03920. Michael Wooldridge and Tomasz Michalak were also supported by the European Research Council under Advanced Grant 291528 (``RACE'').

\bibliographystyle{splncsnat}
\bibliography{references}

\section*{Appendix}
Additionally, we study the performance of our algorithms for the following graphs:
\begin{itemize}
\item \textbf{Cycle:} a cycle of $n$ vertices. A cycle is a simple example of a symmetric, very sparse graph, where the fraction of connected coalitions is very small.
\item \textbf{\citet{erdos1959random}:} a random graph of $n$ nodes, where the probability of an edge between any two vertices is $\mli{edgeProb}$. The graphs obtained using this method has a symmetric degree distribution and relatively large fraction of coalitions that are connected.
\end{itemize}

\begin{figure}[h]
\centerline{
\begin{subfigure}{0.4\textwidth}
\begin{center}
\includegraphics[width=0.97\linewidth]{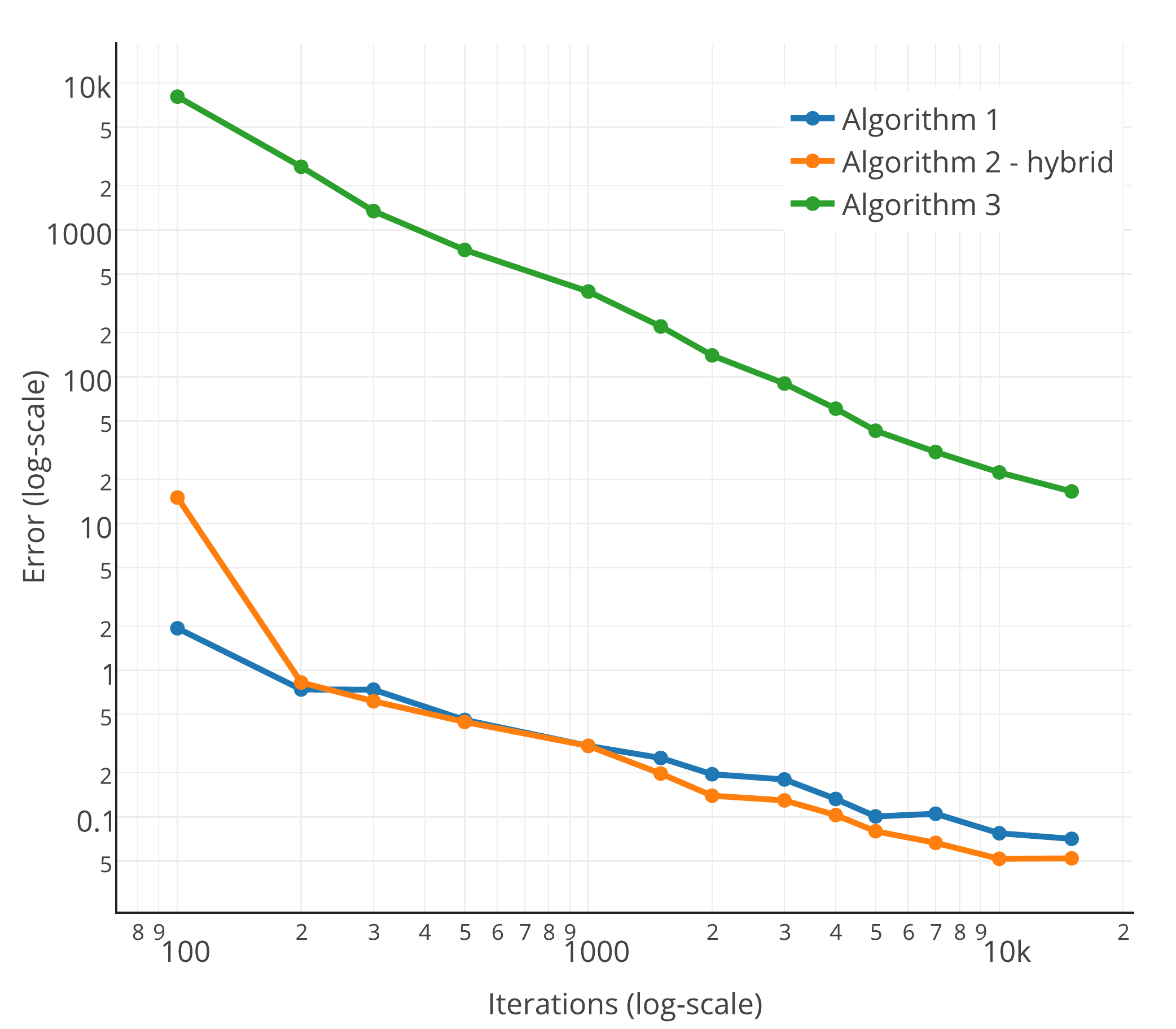}
\end{center}
\caption{Size valuation}
\end{subfigure}
\begin{subfigure}{0.4\textwidth}
\begin{center}
\includegraphics[width=0.97\linewidth]{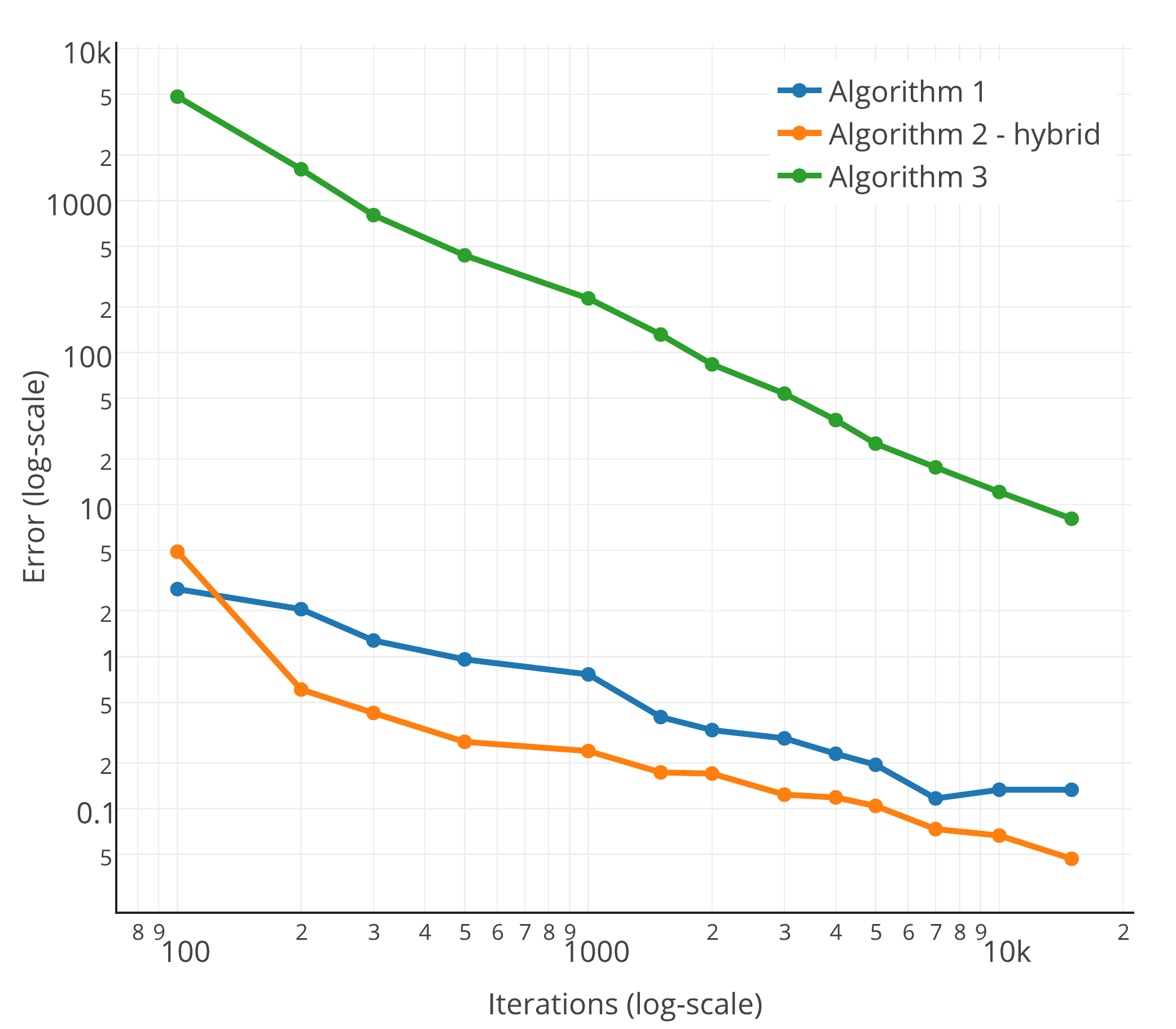}
\end{center}
\caption{Uniform random}
\end{subfigure}
\begin{subfigure}{0.4\textwidth}
\begin{center}
\includegraphics[width=0.97\linewidth]{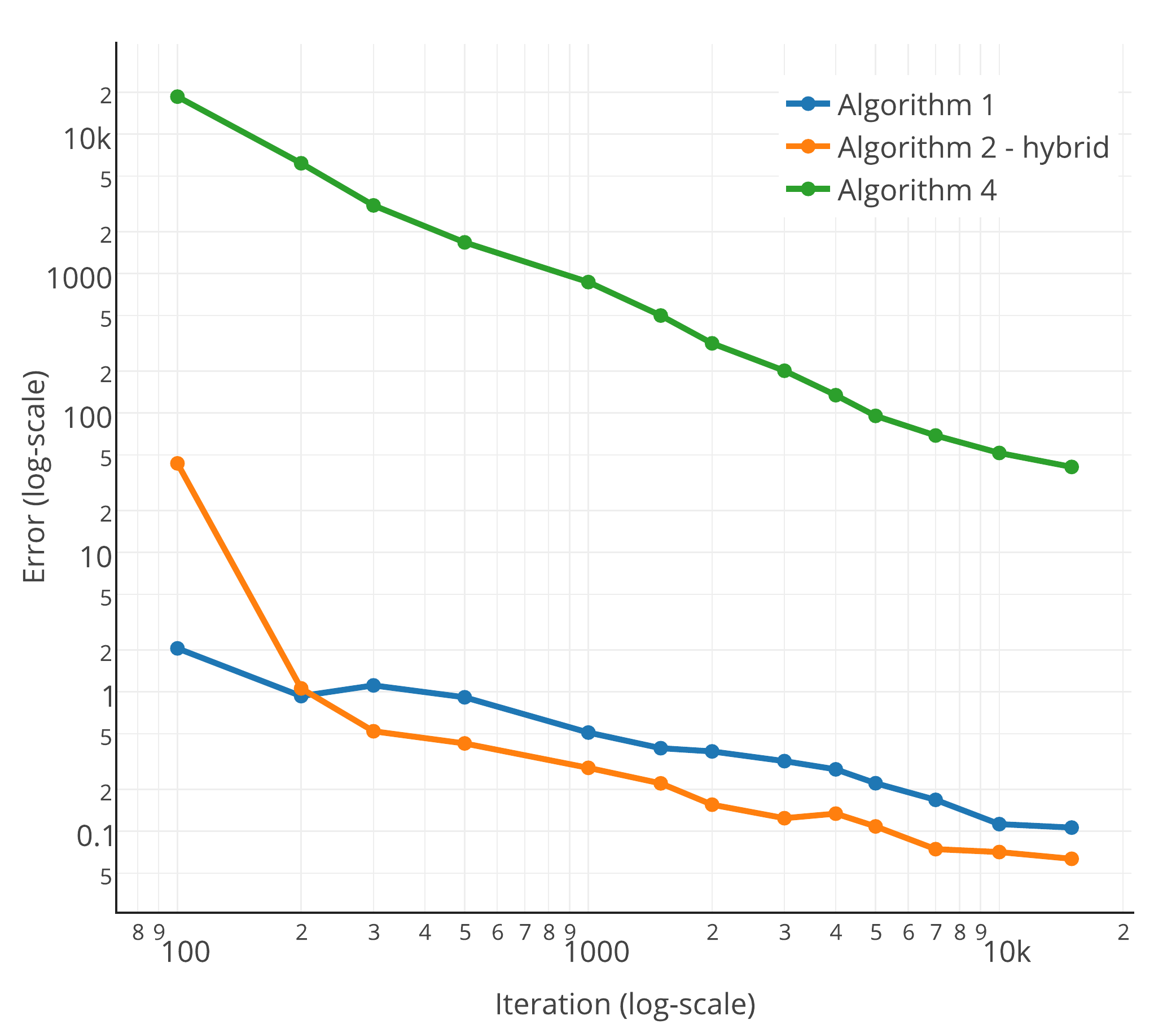}
\end{center}
\caption{Uniform superadditive}
\end{subfigure}
}
\caption{Estimation error against the number of iterations for a cycle graph.}
\label{figure:cycle}
\end{figure}

\begin{figure}
\centerline{
\begin{subfigure}{0.4\textwidth}
\begin{center}
\includegraphics[width=0.97\linewidth]{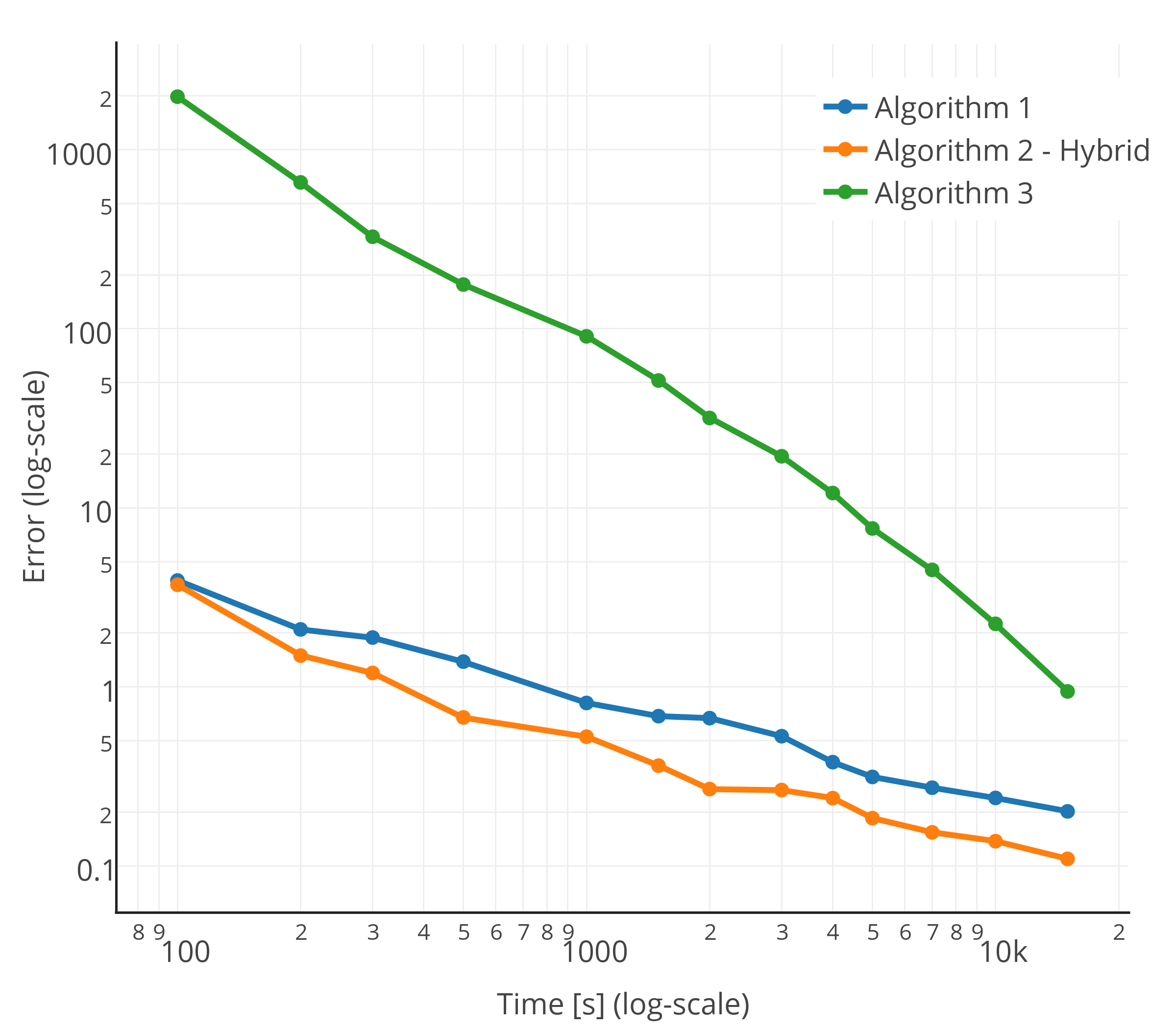}
\end{center}
\caption{Uniform random}
\end{subfigure}
\begin{subfigure}{0.4\textwidth}
\begin{center}
\includegraphics[width=0.97\linewidth]{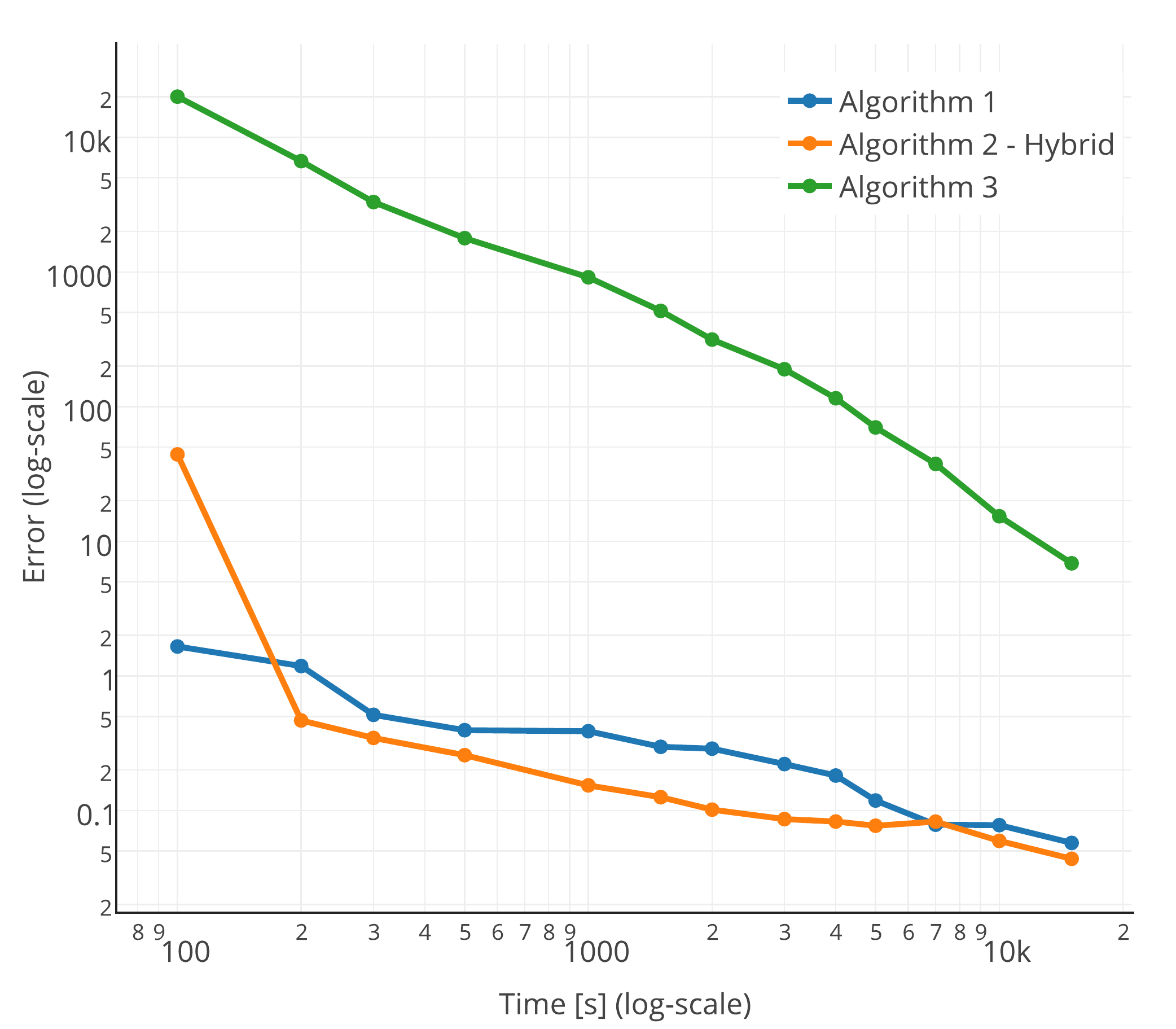}
\end{center}
\caption{Superadditive random}
\end{subfigure}
\begin{subfigure}{0.4\textwidth}
\begin{center}
\includegraphics[width=0.97\linewidth]{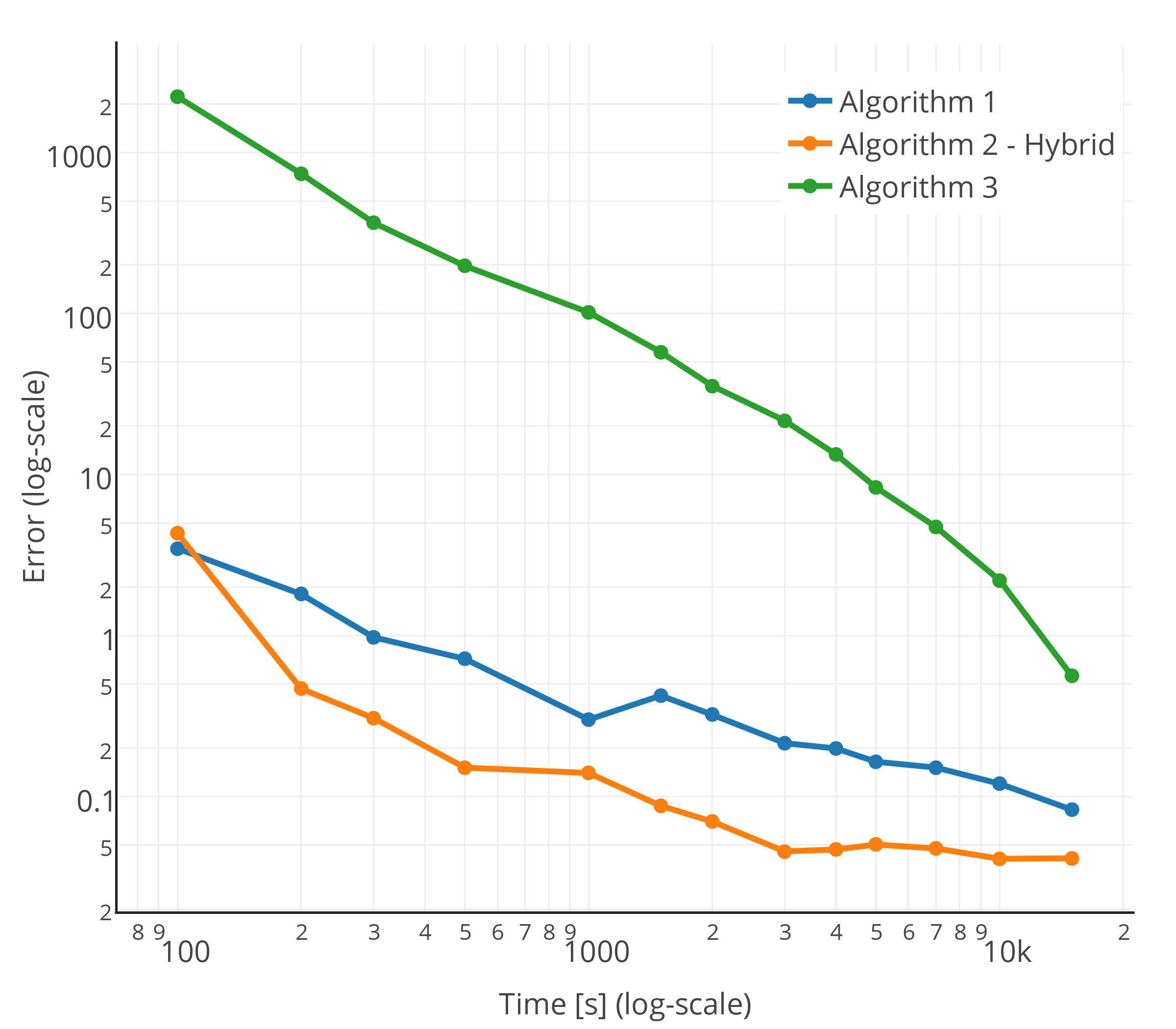}
\end{center}
\caption{Submodular random}
\end{subfigure}
}
\caption{Estimation error against running time for Erdos-Renyi graph with $\mli{edgeProb}=0.4$.}
\end{figure}

\end{document}